\documentclass{article}  
\usepackage{bigsky2009}
\usepackage{graphicx}
\frompage{000} \topage{000}                                              %|
\def\rightmark{Physics with the ALICE EMCal}
\def\leftmark{R. Bellwied et al.}
 
\title{Physics with the ALICE Electromagnetic Calorimeter} 
\authors{
{Rene Bellwied$^1$ for the ALICE Collaboration%
}\\[2.812mm]
{\normalsize
\hspace*{-8pt}$^1$ Physics Department, Wayne State University, \\ 
666 West Hancock, Detroit, MI 48201, USA\\
e-mail: bellwied@physics.wayne.edu\\[0.2ex] 
}}

\abstract{I will present physics measurements which are achievable in the ALICE
experiment at the LHC through the inclusion of a new electromagnetic calorimeter. I will
focus on jet measurements in proton proton and heavy ion collisions. Detailed
simulations have been performed on jet reconstruction, jet
triggering, heavy flavor jet reconstruction through electron identification, gamma-jet
reconstruction and the measurements of identified hadrons and resonances in jets. I will
show the physics capabilities which are made possible through the combination of
calorimeter information with the other detector components in ALICE.}


\begin{document}

\maketitle \setcounter{page}{1}

\section{Introduction}\label{intro}
The addition of an electromagnetic calorimeter to the RHIC experiments (STAR and PHENIX)
has proven to be invaluable for the analysis of neutral pions, photons, electrons, and jets in proton-proton and heavy ion
collisions. Both experiments have recently embarked on full jet reconstruction in CuCu
and AuAu collisions \cite{ploskon,bruna,hei} by combining the charged track information
from their main tracking detectors with the neutral energy information from the
calorimeters.  Past analyses of the interactions of a fast parton with the partonic
medium were based on either single particle or two-particle correlation measurements
which unavoidably carried a strong geometric bias. Models that had very different initial
conditions and dynamic evolution patterns could describe the data, which led to large
ambiguities on the deduced medium properties and the partonic energy loss mechanisms. Only
full jet reconstruction in the heavy ion environment will enable us to fully understand
and quantitatively describe the underlying mechanisms of the interactions between
partons and the hot dense medium. In the following I will first show which physics
questions can be addressed using the ALICE calorimeter information. Then I will describe
recent developments in modeling and analyzing the parton-medium interaction, which are
used to simulate the anticipated physics performance of ALICE as shown
in the concluding chapter. I will close with a brief outlook on the future of in-medium
jet physics at the LHC.

\section{Calorimeter driven physics goals}\label{techno}

\begin{figure}[htb]
\begin{center}
\vspace*{-.1cm}
                 \includegraphics[width=0.6\textwidth]{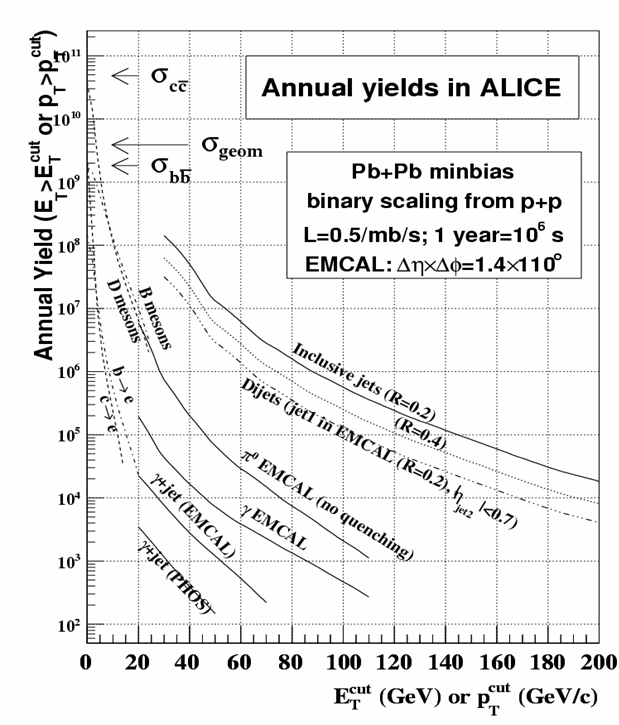}
\vspace*{-0.5cm}
\end{center}
\caption[]{Estimate for triggered jet yields per year in the ALICE EMCal.}
\label{fig1}
\end{figure}

The inclusion of a high resolution energy measurement in ALICE enables us to
determine jet energies in the range shown in Figure \ref{fig1} . Together with the
excellent charged particle reconstruction efficiency in the ALICE tracking system (TPC
plus Inner Tracking System (ITS)) the jet energy measurement accuracy exceeds 80\%. This makes
full jet reconstruction possible. Thus the calorimeter driven physics goals focus
largely on jet related measurements.

Besides the standard hadronic jets the EMCal also adds substantial capability to the
reconstruction of high momentum electrons (from semi-leptonic decay of heavy flavor
jets) and photons from photon jets. Both, electron and photon jets, are an excellent
probe for quark jets, in an energy domain where the hadronic jets are predominantly
formed by gluon fragmentation. I will discuss the background issues for electron and
photon jets in chapter 4. Photon jets are considered 'the golden channel' of jet
reconstruction simply because the photon is not likely to interact and thus carries the
full jet energy. The actual jet measurements (energy loss, medium modification of the
fragmentation) are then performed on the hadronic away-side jet and scaled to the jet energy
information from the reconstructed same side photon. The main measurements that have
been proposed in order to quantify the hot medium effect on the hadronization
are the measurements of the jet yield, the jet suppression factor, the modified fragmentation function, the modified jet shapes, and the sub-jet distributions \cite{vitev,renk,wiedemann}. The inclusion of the
excellent particle identification capabilities of ALICE even at very high momentum
through the relativistic dE/dx measurements in the TPC allow us to measure the
hadro-chemistry of the jets, which has also been proposed as a sensitive measure to
distinguish different parton energy loss models \cite{wiedemann2}. In addition the high
momentum hadrons might be sensitive to flavor conversion mechanisms in the hot,dense
medium \cite{fries}. Finally, the modification of high momentum hadronic resonances in
jets has been proposed as a possible signature for chiral symmetry restoration
\cite{markert}.

\section{New tools for jet physics analysis}

Recent developments in the modeling and reconstruction of jets in high multiplicity
environments, in pp (pile-up) or heavy ion collisions, have
enabled a much more targeted approach to the measurements proposed in the preceding
paragraph. Recently both STAR and PHENIX at RHIC have presented early results of full
jet reconstruction in heavy ion collisions based on these new methods
\cite{ploskon,bruna,hei}.

\subsection{Realistic modeling of parton energy loss in the medium}

The first major advance is based on a series of new Monte Carlo event generators
which try to quantitatively incorporate an in-medium quenching mechanism on the partonic level. The
most advanced are qPYTHIA \cite{salgado}, JEWEL \cite{wiedemann3}, and YaJem
\cite{renk2}. Broadly speaking these models apply an enhanced gluon splitting
probability in the dense partonic medium, which leads to the anticipated changes in the
fragmentation function, i.e. a quenching of the high z (fractional momentum) component
and an increase in multiplicity of low z particles.  The main effect, as suggested by Borghini and Wiedemann \cite{borghini}, is a modification of the so-called hump-back plateau in the vacuum fragmentation function measured  e$^{+}$e$^{-}$  collisions. All these models, which differ considerably in the actual implementation of the quenching effect, are now available from the 
authors.

\subsection{Novel approaches to jet reconstruction}

It was recently realized that the existing cone based jet algorithms, which require
a high momentum particle as a seed, are by definition not infrared and collinear safe
and thus not suitable for quenching studies \cite{salur2,putschke2}. In response the group of
Cacciari, Salam and Soyez has put together the FastJet suite which combines
different new algorithms that have been tested for IR and collinear safety
\cite{salam,soyez,cacciari}. In general the codes can be divided into so-called cone algorithms
(seedless cone, SIS) and recombination algorithms (kT, anti-kT, Cambridge/Aachen cluster).
Recombination algorithms combine (cluster) particles in a 'distant of closest' approach
method and start either with the lowest momentum particles (kT) or the highest momentum
particles (anti-kT). Thus the resulting shape is not necessarily a symmetric cone in
$\Delta\phi$ and $\Delta\eta$. The cone radius parameter has been replaced by a
so-called resolution parameter which limits the extend of clustering based on kinematic
variables. One of the key investigations at this point, which is shared by all three
major LHC experiments, is the relative performance of these jet reconstruction
algorithms for varying multiplicities in pp and heavy ion collisions.

\subsection{Lessons from RHIC analyses}

The RHIC analyses shown at QM09 have revealed a wealth of information in particular
regarding the problems in background subtraction for jet analyses in heavy ion
collisions. The main conclusion was that a data driven approach, i.e. using the detector
response outside of the jet area in order to correct for the background level inside the
cone, proved to be most successful. An additional complication is the occurrence of fake
jets in the sample and alternate solutions to this problem have been proposed
\cite{ploskon,lai,grau}. The jet reconstruction efficiency in heavy ion collisions is
presently deteriorating for jet areas exceeding R=0.4 and jet energies below 20 GeV.
Efforts are underway to further optimize the reconstruction.

\section{Anticipated Physics Performance of the ALICE EMCal}

\subsection{Jet energies and fragmentation functions}

\begin{figure}[htb]
\begin{minipage}{15pc}
\begin{center}
\includegraphics[width=2.7in] {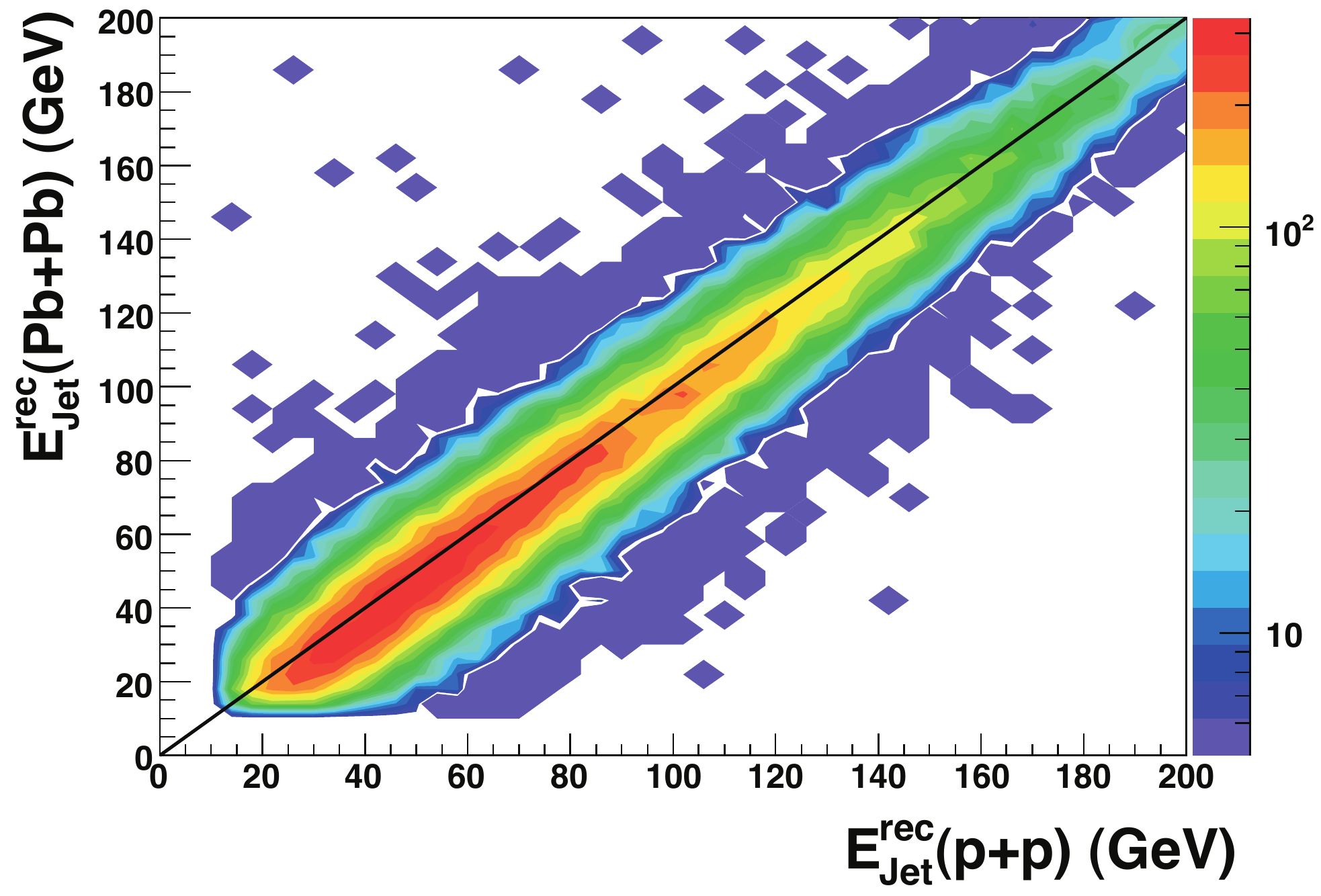}
\end{center}
\end{minipage}\hspace{1pc}%
\begin{minipage}{15pc}
\begin{center}
\includegraphics[width=2.5in] {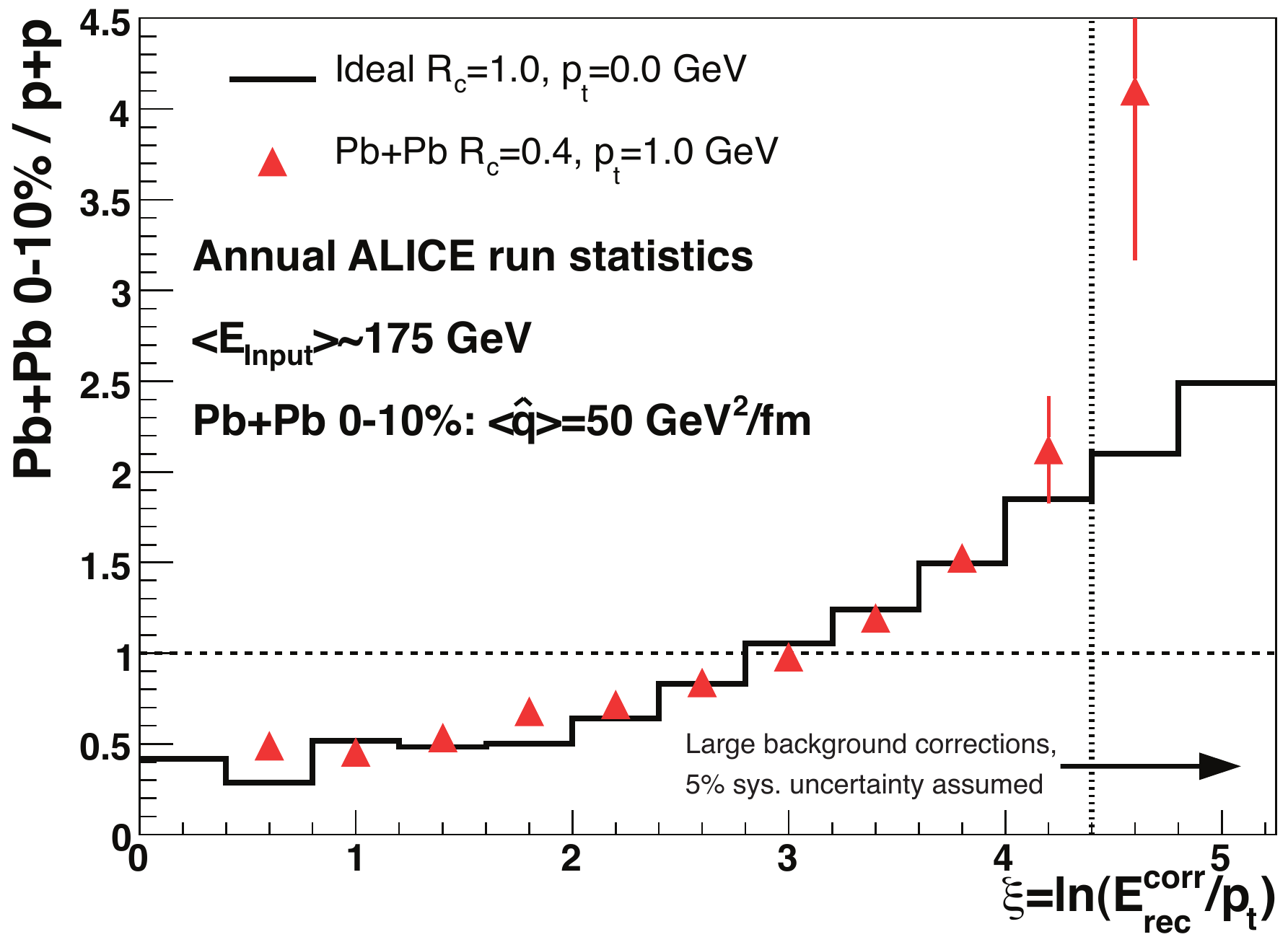}
\end{center}
\end{minipage}
\caption[]{a.) comparison of jet energy resolution in EMCal in pp and AA collisions,
b.) ratio of quenched to unquenched fragmentation function for a 175 GeV jet in the ALICE EMCal reconstructed in pp and PbPb collisions, respectively. The ratio is shown as a function of the inverse fractional momentum. For the quenched simulation a qhat = 50 GeV$^{2}$/fm was assumed. }
\label{fig2}
\end{figure}

Based on our RHIC studies and LHC simulations we expect more than 80\% of the jet
energy to be contained in a R=0.4 area for jet energies higher than 25 GeV. The EMCal 
provides a 15-20\% resolution for these rather low energy jets. In order to conservatively assess our reconstruction capabilities in the heavy ion background, though, we presently limit our
fragmentation and jet yield studies to jet energies of 100 GeV and higher, see Figure
\ref{fig2}a. Through usage of the the new jet reconstruction algorithms we hope to push
this threshold down to 50 GeV in the future. Figure \ref{fig2}b shows the simulated ratio of the quenched
fragmentation function in PbPb collisions to the unquenched function from pp.
This is for a 175 GeV jet assuming a jet modification based on a transport coefficient
of qhat = 50 GeV$^{2}$/fm, which was suggested in the ASW quenching framework
\cite{asw}. More realistic simulations based on the new quenching generators are under
way. As can be seen in the figure we expect to reliably determine the humpback plateau
out to an inverse fractional momentum of about 4.4, which corresponds to a single track
transverse momentum of about 1.5 GeV/c. Ongoing improvements on the hadron and electron
corrections to the jet energy as well as subtracting the effects of elliptic and radial flow will
further reduce the errors on the fractional momentum and thus push the measurement to
higher $\xi$.

\vspace{1.cm}

\subsection{Jet triggering}

\begin{figure}[htb]
\begin{minipage}{15pc}
\begin{center}
\includegraphics[width=2.5in]{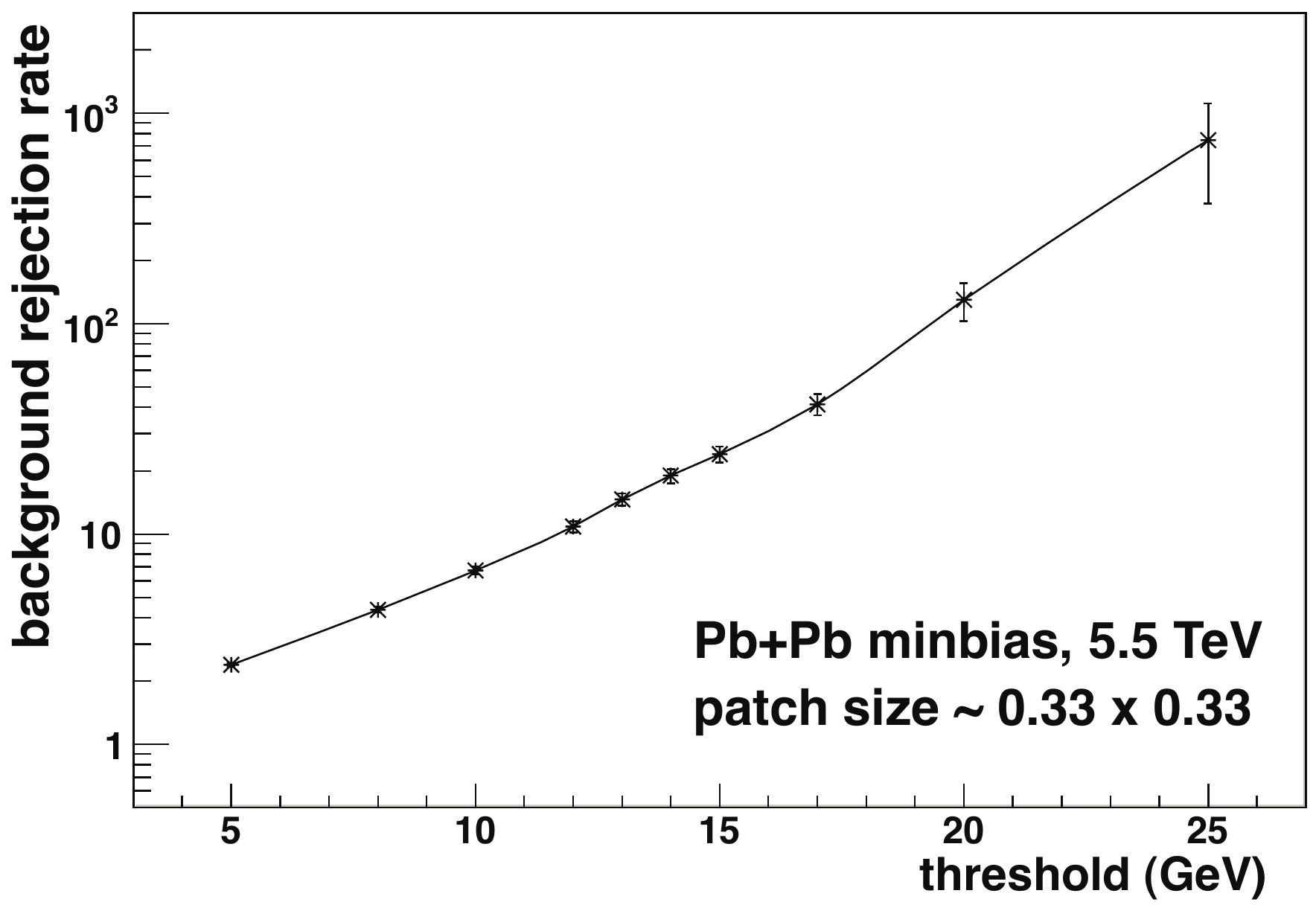} \label{fig3a}
\end{center}
\end{minipage}\hspace{1pc}%
\begin{minipage}{15pc}
\begin{center}
\includegraphics[width=2.5in]{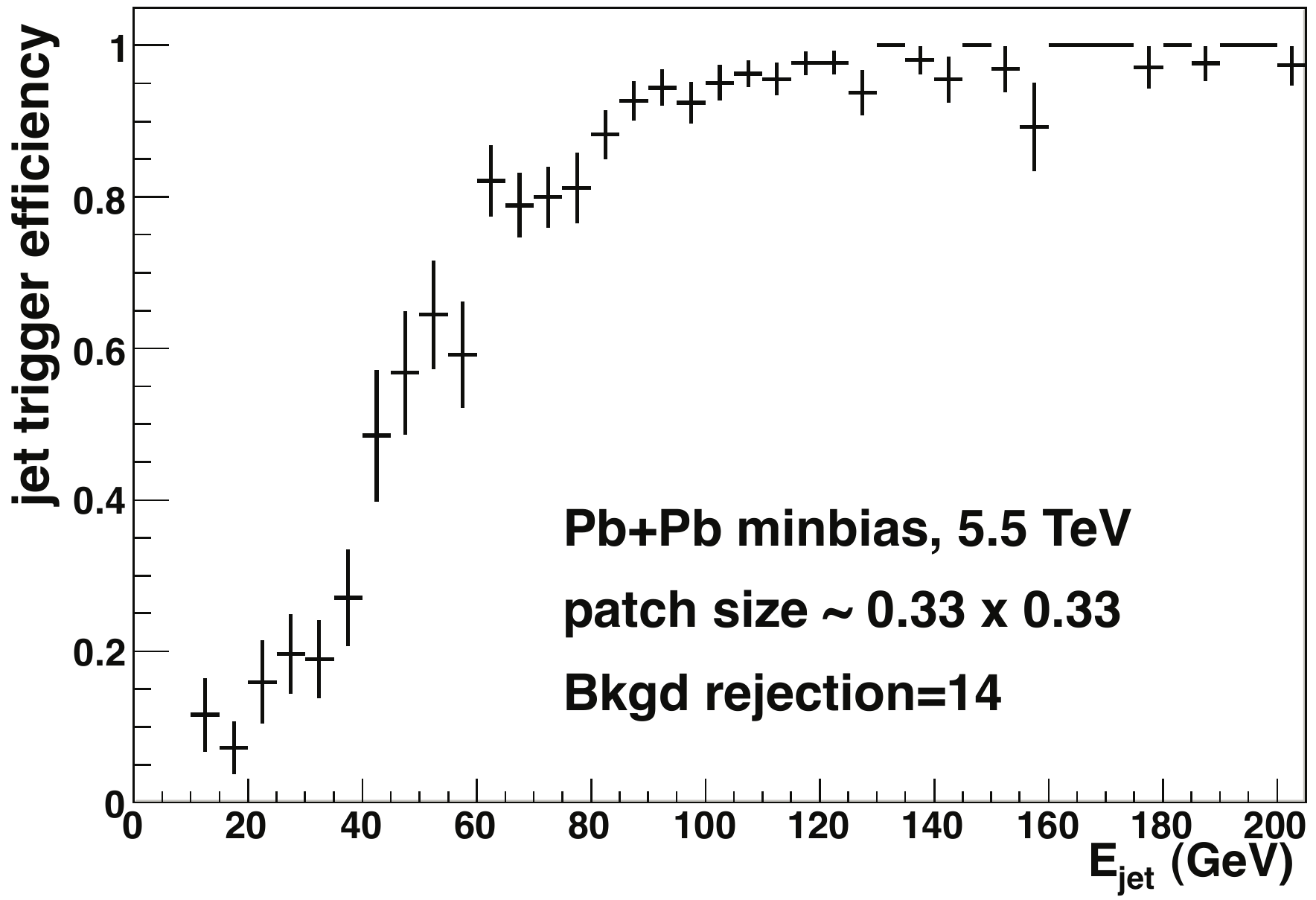}  \label{fig3b}
\end{center}
\end{minipage}
\caption[]{Background rejection rate and L1 jet trigger efficiency based purely on EMCal patch energy information.}
\label{fig3}
\end{figure}

The trigger hierarchy in ALICE allows us to utilize information from the EMCal at two
levels, Level-1 (L1) and the high level trigger (HLT). The L1 trigger is formed after
6.5 $\mu$s and is based on EMCal patch energy information. The resulting jet trigger
efficiency and background rejection rate in Pb-Pb collisions are shown in Figure \ref{fig3}.

The HLT will combine information from the calorimeter, the tracking detector and the
transition radiation detector in order to cut more specifically on photons, electrons
and heavy flavor mesons at high momentum. It will also improve the jet triggering
efficiency down to lower jet energies (as low as 50 GeV). Based on the L1 performance we
already expect an improvement in the jet rate above 100 GeV of a factor 5 in central
PbPb collisions, a factor 50 in peripheral PbPb collisions, and a factor 500 in pp
collisions. Annually we will record around 200,000 jets above a jet energy of 100 GeV in
minimum bias PbPb collisions. Due to the effective jet triggering the identified
particle spectra will reach out to 25 GeV/c for $\pi$,K,p and out to 20 GeV/c for
hadronic resonances (K*, $\Delta$, $\phi$, $\Sigma$*, $\Lambda$*).

\vspace{1.cm}

\subsection{Photon and electron jets}
\begin{figure}[t]
\begin{center}
\vspace*{-.1cm}
                 \includegraphics[width=0.8\textwidth]{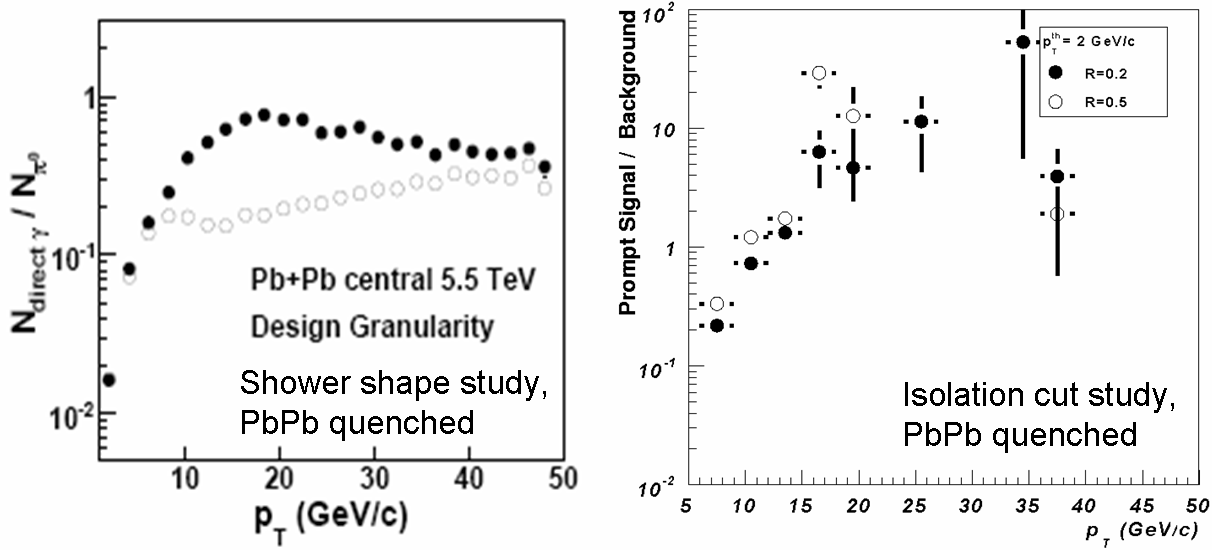}
\vspace*{-0.5cm}
\end{center}
\caption[]{a.) direct gamma to $\pi^{0}$ rate before (open) and after (solid) shower shape analysis, b.) using the output of the shower shape analysis before (open) and after (solid) additional isolation cut.}
\label{fig4}
\end{figure}

The limited coverage in the EMCal allows us to reconstruct photon jets with reasonable
statistics out to a photon energy of about 50 GeV. The main issue with isolating the
golden channel for jet energy measurements is the suppression of background from $\pi^{0}$
decays and from fragmentation photons. Two methods have been applied successively in the
present simulations, a shower shape condition and an isolation cut. The shower shape cut
is based on the fact that a $\pi^{0}$ decay will cause an asymmetric photon shower in
the calorimeter whereas the direct photon showers symmetrically. Details about this
analysis can be found in \cite{gustavo}. Figure \ref{fig4}a shows the improvement in the
$\gamma$/$\pi^{0}$ when the shower shape condition is applied. In the key region of the
photon energy spectrum (20-30 GeV) the ratio reaches unity with the cut. This
sample is then subjected to an isolation  cut based on the energy in the surrounding
calorimeter towers. Another factor 10 is gained (see Figure \ref{fig4}b) in central PbPb
collisions where the pions are assumed to be quenched according to a scaling based on
the measurements at RHIC \cite{phenix-photons}. 

\begin{figure}[h]
\begin{center}
\vspace*{-.1cm}
                 \includegraphics[width=0.5\textwidth]{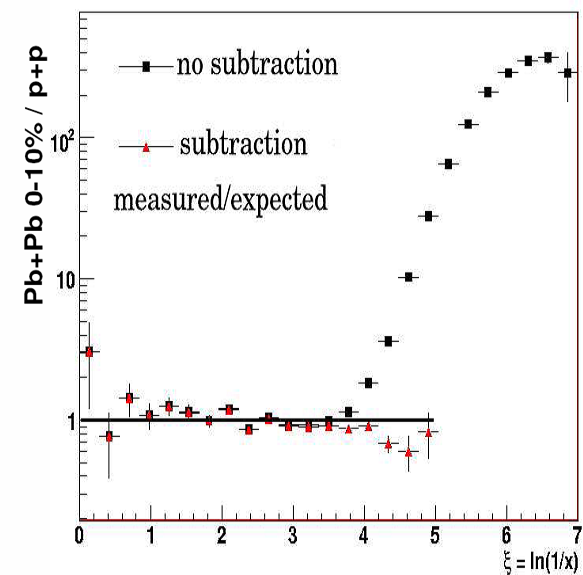}
\vspace*{-0.5cm}
\end{center}
\caption[]{Ratio of quenched to unquenched fragmentation function for a 30 GeV photon jet in the ALICE EMCal reconstructed in pp and PbPb collisions, respectively. The ratio is shown as a function of the inverse fractional momentum. For the quenched simulation a qhat = 50 GeV$^{2}$/fm was assumed. The subtraction curve refers to the result in which the soft bulk matter background, as obtained in Pb-Pb events outside the jet cone, has been subtracted from the Pb-Pb fragmentation function.}
\label{fig5}
\end{figure}

A preliminary study shows that the fragmentation function in the away-side jet for
isolated 30 GeV photon jets can be measured out to an inverse fractional momentum of
about 3.2 as shown in Figure \ref{fig5}.

\clearpage

\begin{figure}[hbtl]
\begin{center}
\vspace*{-.1cm}
                 \includegraphics[width=0.6\textwidth]{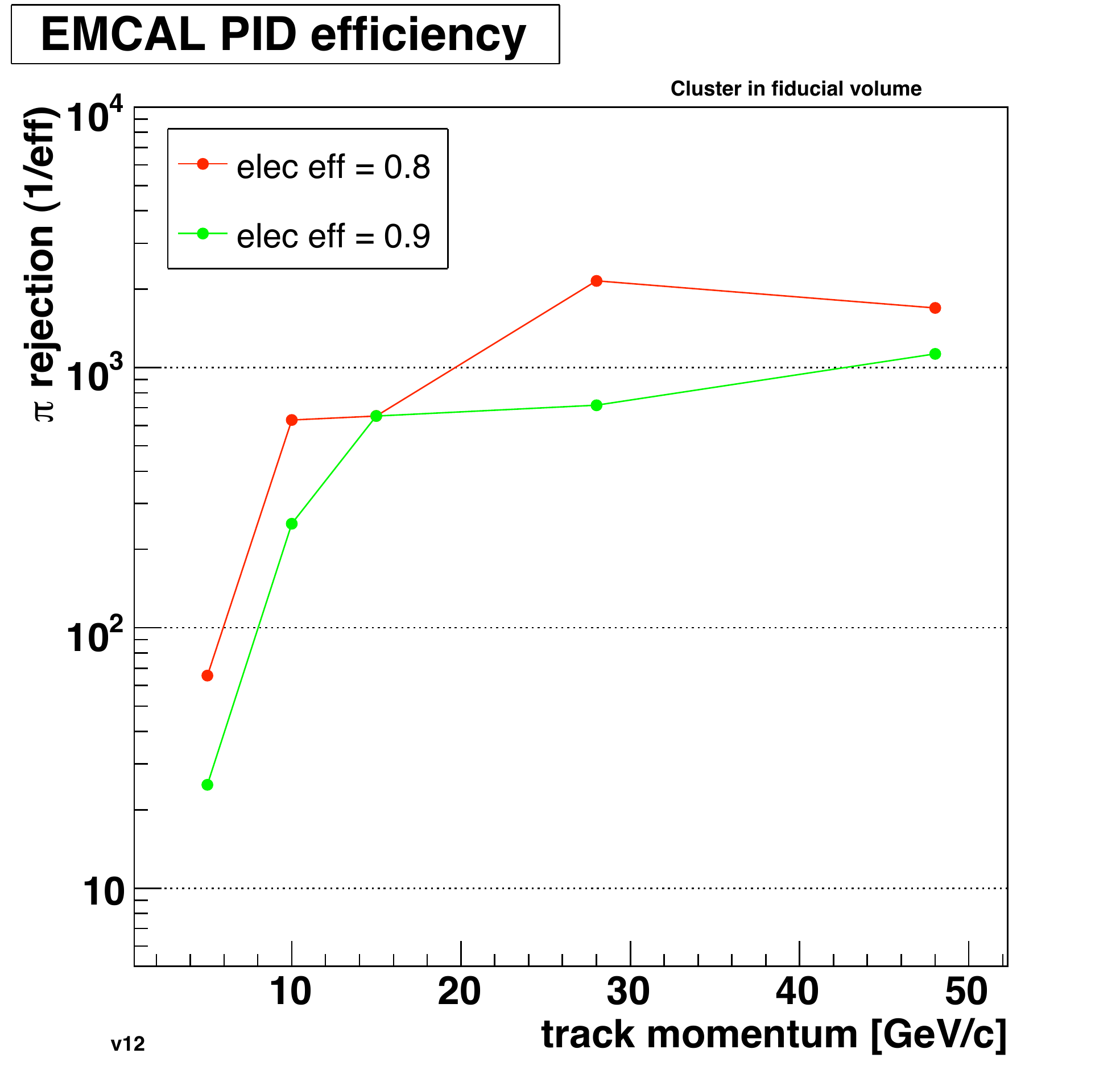}                 
\vspace*{-0.5cm}
\end{center}
\caption[]{Hadron (pion) rejection rate in the EMCal as a function of momentum for several electron reconstruction efficiencies.}
\label{fig6}
\end{figure}

Electron jets are used for heavy quark tagging for two specific reasons. First, the
heavy quark jet is predominantly formed by a fragmenting quark rather than a gluon which
is the dominant process in the untagged jet sample. Second the high momentum heavy quark
supposedly undergoes a different jet energy loss mechanism than the light quarks or the
gluons. Dead cone effects \cite{kharzeev} as well as an enhanced probability for
collisional energy loss \cite{djordjevic} should affect the amount of energy a heavy quark loses in the
partonic medium. This flavor sensitivity of the QCD degrees of freedom in the medium
interaction is not represented in the recently applied AdS/CFT approach to energy loss
\cite{gubser}. In a strongly interacting conformal field theory the energy loss becomes
independent of the degree of freedom and the collision energy. Thus a detailed
measurement of the energy dependence of the energy loss for charm and bottom quarks
should enable us to distinguish between pQCD and AdS/CFT \cite{horowitz}. The key issue
in reliably identifying electrons in the calorimeter is  an effective hadron discrimination.
Figure \ref{fig6} shows preliminary results for pion rejection for different electron
efficiencies as a function of charged track momentum.

\subsection{Hadrons and resonances in jets}

\begin{figure}[hbtl]
\begin{center}
\vspace*{-.1cm}
                 \includegraphics[width=0.8\textwidth]{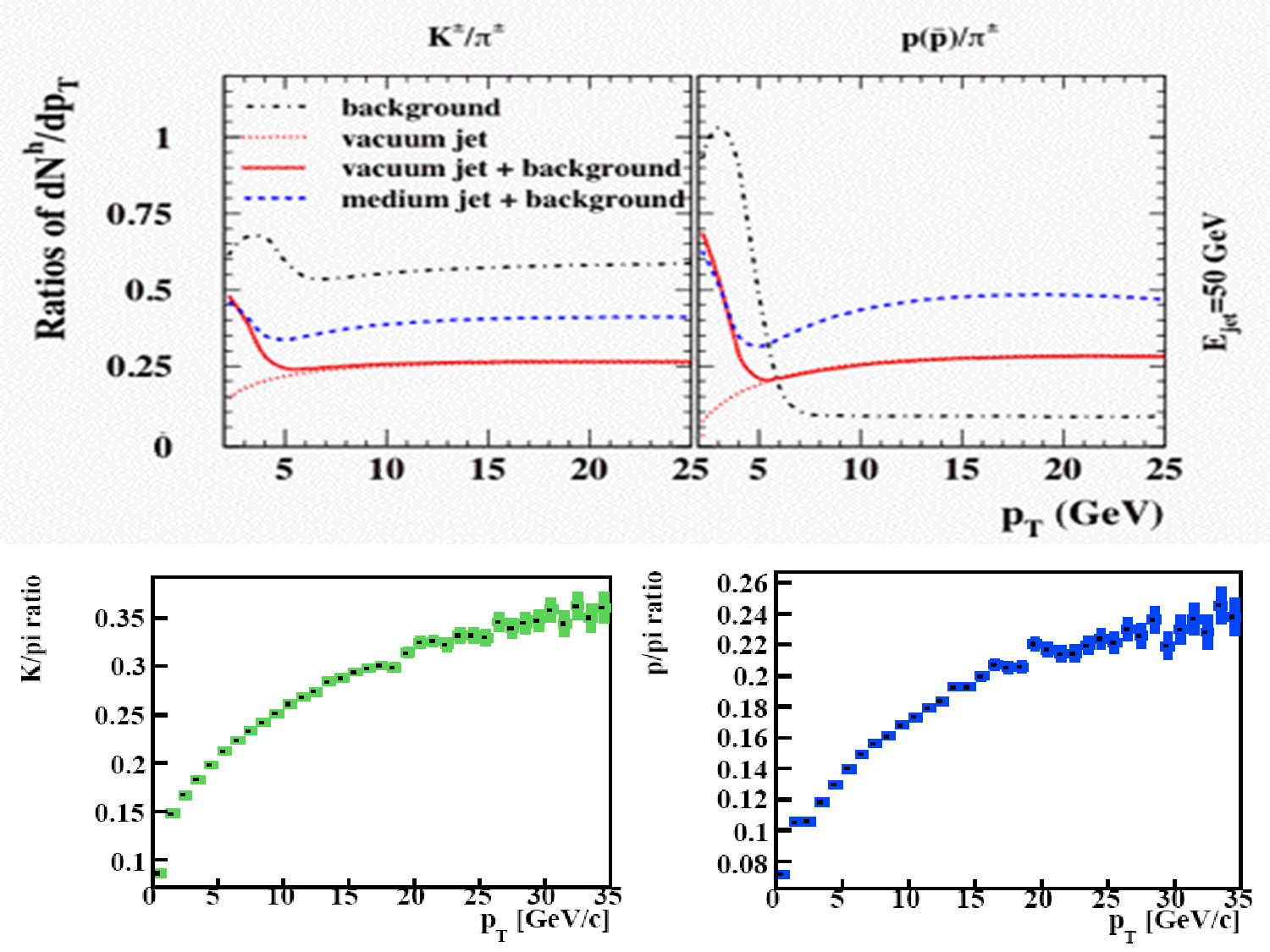}
\vspace*{-0.5cm}
\end{center}
\caption[]{upper row: Prediction by Sapeta and Wiedemann \cite{wiedemann2} on the effect of in-medium gluon splitting on the high momentum hadron ratios, lower row: estimate on the uncertainty in the ratio measurements in ALICE using TPC relativistic dE/dx information on scaled PYTHIA simulations.}
\label{fig7}
\end{figure}

Recent interest in the hadro-chemistry of jets in the medium was triggered by several
theoretical papers. The gluon splitting mechanism, which is the basis of the energy loss
modeling in qPYTHIA and JEWEL, was predicted to also affect the hadro-chemistry of the
quenched jet. Sapeta and Wiedemann postulated that the splitting causes a hadron mass
dependent shift in the particle abundance at high momentum \cite{wiedemann2}. The quenching
will affect the production of higher mass particles less, i.e. ratios such as K/$\pi$ or
p/$\pi$ will increase in the medium compared to fragmentation in vacuum. Figure
\ref{fig7}(upper row) shows a prediction by Sapeta and Wiedemann. The goal of our simulation
was to show that the in-jet particle identification capabilities are such that this
effect could be cleanly measured. Particle identification in ALICE at momenta higher
than 6 GeV/c is presently solely possible through relativistic dE/dx measurements in
the TPC. This method was successfully applied in STAR \cite{lfspectra} and has been
detailed in two technical publications \cite{rdedx1,rdedx2}. Early calibration runs
indicate that the dE/dx resolution of the ALICE-TPC is slightly better than the
STAR-TPC, and based on these cosmic ray results, we have performed a simulation on
the expected PID resolution in the high momentum region. Figure \ref{fig7}(lower row) shows the
result, based on scaled PYTHIA yields, which can be compared directly to the predictions in Figure \ref{fig7}(upper row). 

Clearly ALICE will have the
capability to distinguish between medium and vacuum jets based on particle ratios out to
25 GeV/c. This PID resolution will also enable us to measure potential flavor conversion
effects \cite{fries} which were proposed as an explanation for the nuclear suppression
factor differences for pion, Kaons and protons at high pT as seen in preliminary STAR
data \cite{ant}.

\begin{figure}[hbtl]
\begin{center}
\vspace*{-.1cm}
                 \includegraphics[width=0.5\textwidth] {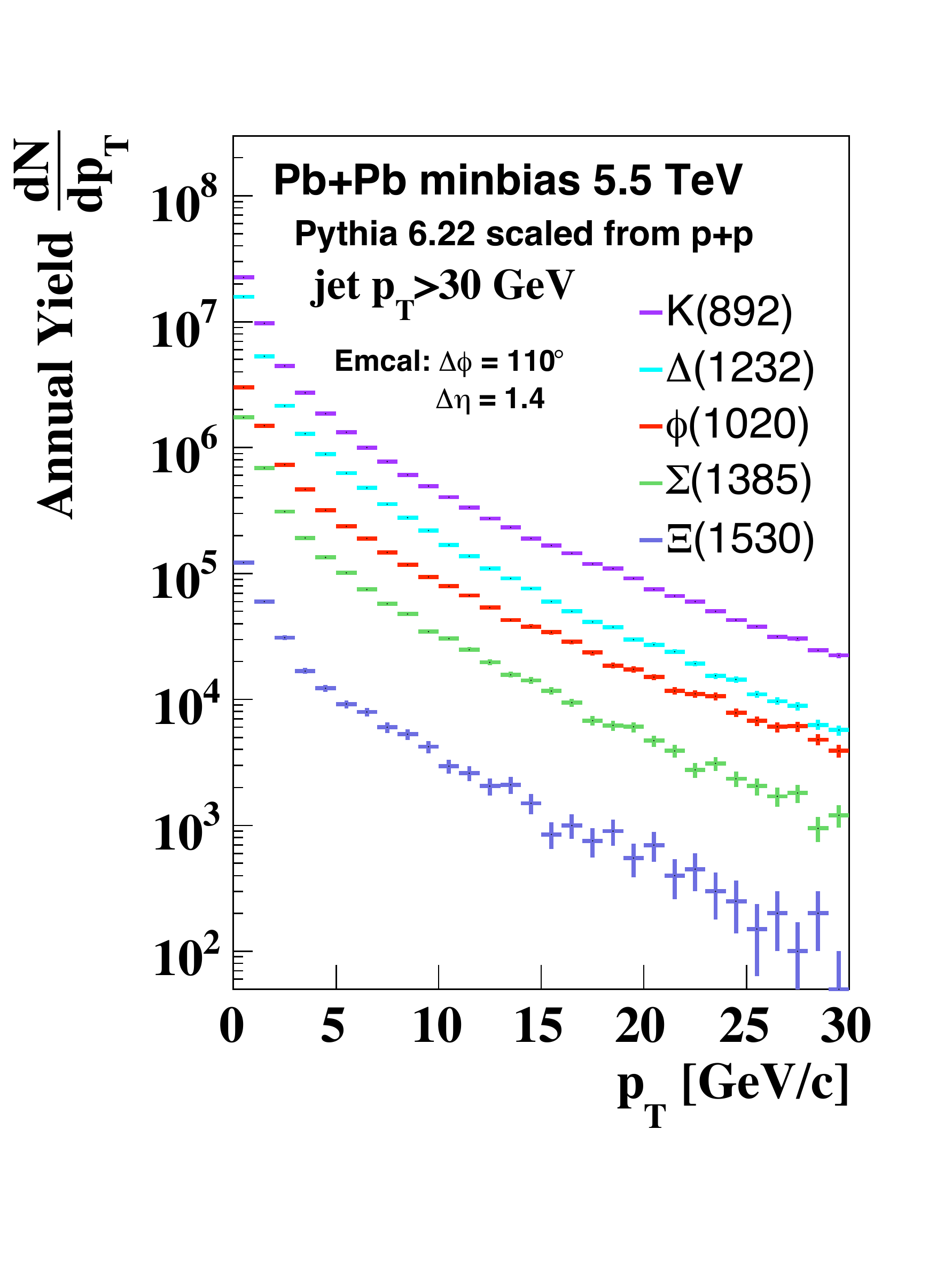}
\vspace*{-1.5cm}
\end{center}
\caption[]{Estimate for yearly reconstructed hadronic resonance yields in jets triggered with the EMCal in PbPb collisions.}
\label{fig8}
\end{figure}

Finally we have embarked on a study of high momentum hadronic resonances in jets. Very
little is known on resonances formation in jets, but a recent paper \cite{markert}
claims that hadronic resonances in a particular high momentum window could actually form
and decay inside the partonic medium. The argument is based on formation time
calculations of pre-hadron (color neutral) states which have been successfully applied
to explain hadron attenuation measurements in cold nuclear matter \cite{hermes,accardi}.
If the resonances can indeed form and decay off-shell in the medium, then they would be
a probe sensitive to chiral symmetry restoration. The proposed signal
\cite{markert} will require to measure resonances in quadrants relative to a jet axis
in order to distinguish between jet and bulk resonances. Figure \ref{fig8} shows the
available statistics for such a measurement. An early feasibility test was performed on
STAR data for lower momentum resonances using the highest pT particle in the event rather
than the reconstructed jet axis  \cite{markert2}. This analysis will benefit tremendously from the 
increased statistics and jet reconstruction capability in ALICE.  In addition the reconstruction of resonance over non-resonant hadrons (e.g K*/K, $\Lambda$*/$\Lambda$) ratios has been used
previously \cite{markert3} in conjunction with HBT data to determine the partonic and
hadronic lifetimes of the medium formed in heavy ion collisions.

\clearpage

\section{Outlook}

The heavy ion program at the LHC benefits tremendously from the inclusion of an
electromagnetic calorimeter in ALICE. By bringing extended jet reconstruction
capabilities to the full suite of ALICE sub-systems, which have been optimized for
relativistic heavy ion research, the experiment has extended its portfolio into a regime
that proved to be very important in the ongoing RHIC experiments. STAR and PHENIX have
been successful in the high pT regime by making first measurement of nuclear
suppression factors, direct photons and semi-leptonic decays of heavy mesons, but they
lack the statistics to unambiguously reconstruct jets at sufficiently high
momentum to embark on a quantitative and systematic exploration of the energy loss and
medium effects of partons in the hot dense matter. Although the ALICE EMCal lacks the
coverage of the comparable calorimeters in CMS and ATLAS its pairing with the superior
tracking and particle identification detectors in ALICE allows for a unique set of pp
and heavy ion measurements. In particular the particle identified measurements in jets,
as well as the correlation of identified photons and electrons with jets and jet
particles down to very low fractional momentum are unique to ALICE. We expect to
quantitatively solve the questions of quark and gluon energy loss as well as medium
response to the jets. The uncertainties of the degrees of freedom above the critical
temperature and chiral symmetry restoration will be addressed as well.

\section*{Acknowledgments}

I would like to thank all my colleagues in ALICE for their input, in particular the
ALICE-U.S. collaboration, John Harris, Peter Jacobs, and Christina Markert.

\vfill\eject
\end{document}